\documentclass{ws-ijmpd}
\usepackage[super]{cite}
\usepackage{graphicx}
\begin{document}

\markboth{H. S. VIEIRA et al.}
{Resonant frequencies of a charged scalar field in the GHS dilaton black hole}

\catchline{}{}{}{}{}

\title{Resonant frequencies of a charged scalar field in the Garfinkle-Horowitz-Strominger dilaton black hole}

\author{H. S. VIEIRA}
\address{Departamento de F\'{i}sica, Universidade Federal da Para\'{i}ba, Caixa Postal 5008, CEP 58051-970, Jo\~{a}o Pessoa, PB, Brazil\\
horacio.santana.vieira@hotmail.com}

\author{V. B. BEZERRA}
\address{Departamento de F\'{i}sica, Universidade Federal da Para\'{i}ba, Caixa Postal 5008, CEP 58051-970, Jo\~{a}o Pessoa, PB, Brazil\\
valdir@fisica.ufpb.br}

\author{C. R. MUNIZ}
\address{Grupo de F\'isica Te\'orica (GFT), Universidade Estadual do Cear\'a, Faculdade de Educa\c c\~ao, Ci\^encias e Letras de Iguatu, Iguatu, Cear\'a, Brazil\\
celio.muniz@uece.br}

\author{M. S. CUNHA}
\address{Grupo de F\'isica Te\'orica (GFT), Centro de Ci\^encias e Tecnologia, Universidade Estadual do Cear\'a, CEP 60714-903, Fortaleza, Cear\'a, Brazil\\
marcony.cunha@uece.br}

\maketitle

\begin{history}
\received{Day Month Year}
\revised{\today}
\end{history}

\begin{abstract}
We use the exact analytical solution of the radial part of the Klein-Gordon equation describing a charged massive scalar field in the electrically charged Garfinkle-Horowitz-Strominger dilaton black hole spacetime, given in terms of the confluent Heun functions, to study the physical phenomena related to resonant frequencies associated to this field, and also to examine some aspects related to its Hawking radiation. The special case of a Schwarzschild spacetime is also considered.

\end{abstract}

\ccode{PACS Nos.: 02.30.Gp, 03.65.Ge, 04.20.Jb, 04.62.+v, 04.70.-s}
%
%
\section{Introduction}
The general theory of relativity predicts the existence of black holes which, initially, were considered as a simple mathematical consequence of Einstein's equations rather than real objects existing in some region of our Universe. Nowadays, with the recent detection of gravitational waves produced in black holes collisions \cite{PhysRevLett.116.221101} and the measurements of the properties of a source in M87 compatible with the presence of a central Kerr black hole \cite{AstrophysJLett.875.L1} confirm that these objects really exist and this fact motivate us to do more investigations in order to understand their physics. On the other hand, from the theoretical point of view, the studies about black holes also will help us to better understand their physics and possibly to use this to formulate a theory reconciling quantum mechanics and gravity.

The spacetime generated by a black hole with large mass as compared to the Planck mass, has small curvature in the region surrounding the horizon and outside it. In these regions, the equations describing gravity in the context of string theory can be approximated by Einstein's equations. Thus, in this scenario, the Schwarzschild solution of general relativity could be a good approximation to describe static and uncharged black holes  in the string theory. But, when we consider the solutions of the Einstein-Maxwell equations, the predictions of the theory differ from the ones obtained in general relativity due to the presence of a scalar field called dilaton, which couples with Maxwell field and changes completely the properties of the black hole. One example which confirms this fact is the solution of Reissner-Nordstr\"{o}m describing a static, charged and spherically-symmetric black hole in general relativity. This spacetime is not a solution of the string theory even approximately, at low energy regime. Thus, the addition of the dilaton field changes the properties of the black holes. The first studies on the solutions of black holes in the context of string theory were performed in the 1980's \cite{NuclPhysB.207.337,AnnPhys.172.304,NuclPhysB.289.701,NuclPhysB.309.552,NuclPhysB.298.741}. In the early 1990's, Garfinkle, Horowitz and Strominger \cite{PhysRevD.43.3140} obtained a solution which corresponds to static, charged and spherically-symmetric black hole in the low energy limit of string theory and Sen \cite{PhysRevLett.69.1006} obtained an asymptotically flat black hole solution in the same limit of the string theory. These black holes solutions in string theory are important due to the fact that this theory is a good candidate to an eventual quantum theory of gravity.

In this paper we will focus on a family of spherically symmetric static and charged black hole solutions obtained in the low energy limit of heterotic string field in (3+1)-dimensions, know as the Garfinkle-Horowitz-Strominger dilaton black hole (GHS dilaton black hole). In this spacetime we will consider a charged and massive scalar field, whose solution of the radial part of the Klein-Gordon equation is given in terms of the confluent Heun functions \cite{CanJPhys.87.349,arXiv:1807.09135v1,AdvHighEnergyPhys.2019.5769564}. We will use this solution to determine the resonant frequencies and to discuss the Hawking radiation of scalar particles.

Other studies were performed in the GHS dilaton black hole spacetime, as for example, on some aspects of thermodynamics \cite{ChinPhysB.19.090401,NuovoCimento.122.904,CommunTheorPhys.52.184}, the quasinormal modes \cite{PhysRevD.63.064009,PhysRevD.70.084046,ClassQuantumGrav.22.1129}, and about the resonance spectra \cite{AdvHighEnergyPhys.2015.739153,PhysRevD.81.104042}.

The resonant frequencies are one of the essential characteristics of a black hole and play an important role with respect to the radiation emitted by black holes. They correspond to damped oscillations and represent a kind of ``sound'' produced by a black hole. Thus, it is possible to get some information about the physics of black holes through the resonant frequencies. For this reason it is important to compute them. As we are considering the interaction between a GHS dilaton black hole and scalar fields, in order to obtain these resonant frequencies we impose that the Heun functions, which are solutions of the radial part of the Klein-Gordon equation, should have a polynomial form.

It is worth emphasizing that the spectra corresponding to these resonant frequencies are related to the decay of the perturbation field, while the quasinormal modes (QNMs) are solutions of the perturbation equations with appropriate boundary conditions which are imposed on the outgoing waves and on waves crossing the horizon \cite{Detweiler:1979,PhysLettB.761.53,EurPhysJPlus.132.324}, and thus, in principle, we can obtain the former ones from the later.

We also use this solution in terms of Heun functions to study the Hawking radiation of scalar particles in the GHS dilaton black hole background spacetime. This radiation \cite{Nature.248.30,CommunMathPhys.43.199} is associated with the interaction of quantum fields and the curvature of the spacetime and therefore it is an interesting semi-classical phenomena which can give us some insights about the physics of a black hole and for this reason should be investigated.

These results, concerning the resonant frequencies \cite{AnnPhys.373.28,PhysRevD.94.084040}, as well as the ones related to the Hawking radiation, are compared with similar results obtained in the Schwarzschild black hole spacetime with the aim to emphasize the role played by the dilaton field.

This paper is organized as follows. In Section 2, we reobtain, for the sake of completeness, the solutions of the Klein-Gordon equation for a charged massive scalar field in the GHS dilaton black hole spacetime. In Section 3, we find the resonant frequencies for both massive and massless scalar particles. In Section 4, we investigate some aspects of the Hawking radiation. Finally, in Section 5, the conclusions are given. The units where $G=c=\hbar=1$ were chosen.
%
%
\section{Klein-Gordon equation in the GHS dilaton black hole}\label{Sec.II}
The geometry of the GHS dilaton black hole is obtained from a (3+1)-dimensional action which describes a system where a dilaton filed, $\phi$, is coupled to electromagnetic field in such a way that the action is given by
\begin{equation}
S=\int d^{4}x \sqrt{-g} [R-2(\nabla\phi)^{2}-\mbox{e}^{-2\phi}F^{2}]\ ,
\label{eq:action_GHS}
\end{equation}
where $F_{\mu\nu}$ is the electromagnetic field tensor.

In this work we consider the solution obtained from this action, in the low energy effective regime in string theory. In this scenario, the metric corresponding to a static, spherically symmetric and charged dilaton black hole, called GHS dilaton black hole, in spherical coordinates, is given by \cite{PhysRevD.43.3140}
\begin{equation}
ds^{2}=-\left(1-\frac{2M}{r}\right)dt^{2}+\left(1-\frac{2M}{r}\right)^{-1}dr^{2}+r(r-a)\ d\Omega^{2}\ ,
\label{eq:metrica_GHS}
\end{equation}
with
\begin{equation}
d\Omega^{2}=d\theta^{2}+\sin^{2}\theta\ d\varphi^{2}\ .
\label{eq:solid_angle}
\end{equation}

This kind of black hole with electric or magnetic charge surrounded by a dilaton field was studied in different contexts, specially in string theories in which case it arises as a solution in the low energy regime of an effective four-dimensional theories.

The parameter $a$ is related to the dilaton field, namely,
\begin{equation}
a=\frac{Q^{2}\mbox{e}^{-2\phi_{0}}}{M}\ ,
\label{eq:parameter_a_GHS}
\end{equation}
where $\phi_{0}$ is the asymptotic value of the dilaton field, such that
\begin{equation}
\mbox{e}^{-2\phi}=\mbox{e}^{-2\phi_{0}}\biggl(1-\frac{Q^{2}\mbox{e}^{-2\phi_{0}}}{Mr}\biggr)\ ,
\label{eq:phi_GHS}
\end{equation}
with $M$ and $Q$ being the physical mass (total mass) and the magnetic or electric charge of the GHS dilaton black hole, respectively. The metric given by Eq.~(\ref{eq:metrica_GHS}) is similar to the Schwarzschild metric with a difference that the area of the sphere for $t$ and $r$ constants in the GHS dilaton black hole depends on the charge (intensity of the dilaton field). On the other hand, this metric is completely different from the Reissner-Nordstr\"{o}m metric. For our purposes, let us focus on $\phi_{0}=0$. Notice that if we consider $Q$ as the electric charge, when it is equal to zero, $Q=0$, which implies $a=0$, we have a limiting case which corresponds to the Schwarzschild spacetime. Thus, the GHS dilaton black hole differs from the Schwarzschild black hole due to the presence of the scalar field called dilaton, which will change the properties of the black hole geometry as compared with the geometry of the Schwarzschild black hole.

Now, let us examine the interaction between a charged scalar fields and the GHS dilaton black hole. In order to do this, we will consider the Klein-Gordon equation, which can be write as
\begin{eqnarray}
&& \biggl[\frac{1}{\sqrt{-g}}\partial_{\sigma}(g^{\sigma\tau}\sqrt{-g}\partial_{\tau})-ie(\partial_{\sigma}A^{\sigma})-2ieA^{\sigma}\partial_{\sigma}-\frac{ie}{\sqrt{-g}}A^{\sigma}(\partial_{\sigma}\sqrt{-g})\nonumber\\
&& -e^{2}A^{\sigma}A_{\sigma}-\mu_{0}^{2}\biggr]\Psi=0\ ,
\label{eq:Klein-Gordon_gauge_GHS}
\end{eqnarray}
with
\begin{equation}
\sqrt{-g}=r(r-a)\sin\theta\ ,
\label{eq:determinant}
\end{equation}
where $\mu_{0}$ is the mass of the scalar particle, and $e$ is its charge. As we are considering that the GHS dilaton black hole is electrically charged, we have that the 4-vector electromagnetic potential is given by \cite{NuclearPhysicsB.899.37}
\begin{equation}
A_{\sigma}dx^{\sigma}=-\frac{Q}{r}\ dt\ .
\label{eq:potencial_EM_GHS}
\end{equation}
Note that the magnetically charged solutions can be obtained by applying a duality transformation on the electromagnetic field and changing the sign of the dilaton field $(\phi \rightarrow -\phi)$. This means that the dilaton electric and magnetic fields have opposite signs, and that the geometry is preserved.

Thus, substituting Eq.~(\ref{eq:metrica_GHS}) into Eq.~(\ref{eq:Klein-Gordon_gauge_GHS}), we obtain
\begin{eqnarray}
&& \biggl\{-\frac{r^{2}(r-r_{d})}{r-r_{h}}\frac{\partial^{2}}{\partial t^{2}}+\frac{\partial}{\partial r}\biggl[(r-r_{h})(r-r_{d})\frac{\partial}{\partial r}\biggr]-\mathbf{L}^{2}\nonumber\\
&& -\frac{2i e Q}{r-r_{h}}r(r-r_{d})\frac{\partial}{\partial t}+\frac{e^{2}Q^{2}(r-r_{a})}{r-r_{h}}-r(r-r_{d})\mu_{0}^{2}\biggr\}\Psi=0\ ,
\label{eq:mov_1_GHS}
\end{eqnarray}
where $r_{h}=2M$, $r_{d}=a$, and $\mathbf{L}^{2}$ is the angular momentum operator given by
\begin{equation}
\mathbf{L}^{2}=-\frac{1}{\sin\theta}\frac{\partial}{\partial \theta}\biggl(\sin\theta\frac{\partial}{\partial \theta}\biggr)-\frac{1}{\sin^{2}\theta}\frac{\partial^{2}}{\partial\varphi^{2}}\ .
\label{eq:angular_operator_GHS}
\end{equation}

Due to the spacetime symmetry we can separate the scalar wave function as
\begin{equation}
\Psi(\mathbf{r},t)=R(r)Y_{l}^{m}(\theta,\varphi)\mbox{e}^{-i \omega t}\ ,
\label{eq:separacao_variaveis_GHS}
\end{equation}
where $Y_{l}^{m}$ are the spherical harmonics, $l=\{0,1,2,...\}$ and $|m| \leq l$ are the orbital and the azimuthal quantum numbers, respectively. The frequency (energy) is taken as $\omega > 0$, which corresponds to the flux of particles at infinity. Therefore, by using this separation of variables, we can write Eq.~(\ref{eq:mov_1_GHS}) as
\begin{eqnarray}
&& \frac{d}{dr}\biggl[(r-r_{h})(r-r_{d})\frac{dR}{dr}\biggr]\nonumber\\
&& +\biggl[\frac{r-r_{d}}{r-r_{h}}(\omega r-eQ)^{2}-\lambda_{lm}-r(r-r_{d})\mu_{0}^{2}\biggr]R=0\ ,
\label{eq:mov_radial_1_GHS}
\end{eqnarray}
where $\lambda_{lm}=l(l+1)$.

The solutions for the angular part are different from the one obtained in the literature \cite{CanJPhys.87.349} which considers a magnetically charged GHS dilaton black hole. Otherwise, in Ref.~\refcite{AdvHighEnergyPhys.2019.5769564} which considers the electrically charged case, the solutions are the same that we obtained in a previous version of the present paper \cite{arXiv:1807.09135v1}.
%
%
\subsection{Radial equation}
Until recently, the solutions of the radial Klein-Gordon equation in the GHS dilaton black hole spacetime were known only in the asymptotic regimes, namely, very close to the event horizon and far from the black hole \cite{ClassQuantumGrav.22.533,AstrophysSpaceSci.333.369}. Otherwise, it is possible to know the solution in all region exterior to the event horizon, but only numerically \cite{PhysRevD.70.084046,IntJTheorPhys.49.2786,IntJTheorPhys.52.1474}. Nowadays, we know an exact solution for a scalar field in the GHS dilaton black hole magnetically charged \cite{CanJPhys.87.349}. In what follows we will reobtain the exact analytical solution of the radial equation, in the case under consideration, namely, a GHS dilaton black hole electrically charged, in the exterior region to the event horizon. It is worth calling attention to the fact that as the geometry of the black hole is preserved independently of the dilaton charge, if electric or magnetic, rigorously speaking, it was not necessary to reobtain the radial solution.

Thus, to solve exactly Eq.~(\ref{eq:mov_radial_1_GHS}) we use the procedure developed in our recent papers (see for example Refs.~\refcite{ClassQuantumGrav.31.045003,AnnPhys.350.14,JCAP01(2018)006} and references therein). Adopting this procedure, the general solution of the radial equation, given by Eq.~(\ref{eq:mov_radial_1_GHS}), in the region exterior to the event horizon, is given by
\begin{eqnarray}
R(x) & = & \mbox{e}^{\frac{1}{2}\alpha x}x^{\frac{1}{2}\beta}\{C_{1}\ \mbox{HeunC}(\alpha,\beta,\gamma,\delta,\eta;x)\nonumber\\
&& +C_{2}\ x^{-\beta}\ \mbox{HeunC}(\alpha,-\beta,\gamma,\delta,\eta;x)\}\ ,
\label{eq:solucao_geral_radial_GHS}
\end{eqnarray}
where
\begin{equation}
x=\frac{r-r_{h}}{r_{d}-r_{h}}\ ,
\label{eq:x_GHS}
\end{equation}
with $C_{1}$ and $C_{2}$ being constants, and the parameters $\alpha$, $\beta$, $\gamma$, $\delta$, and $\eta$ given by
\begin{equation}
\alpha=2(r_{h}-r_{d})(\mu_{0}^{2}-\omega^{2})^{\frac{1}{2}}\ ,
\label{eq:alpha_radial_HeunC_GHS}
\end{equation}
\begin{equation}
\beta=2i(\omega r_{h}-eQ)\ ,
\label{eq:beta_radial_HeunC_GHS}
\end{equation}
\begin{equation}
\gamma=0\ ,
\label{eq:gamma_radial_HeunC_GHS}
\end{equation}
\begin{equation}
\delta=(r_{h}-r_{d})[2eQ\omega+r_{h}(\mu_{0}^{2}-2\omega^{2})]\ ,
\label{eq:delta_radial_HeunC_GHS}
\end{equation}
\begin{equation}
\eta=-\lambda_{lm}-\delta\ .
\label{eq:eta_radial_HeunC_GHS}
\end{equation}

As $\beta$ is not necessarily an integer, the Heun's functions that appear in Eq.~(\ref{eq:solucao_geral_radial_GHS}) are linearly independent solutions of the confluent Heun dif\-fer\-en\-tial equation, which can be written as \cite{JPhysAMathTheor.43.035203}
\begin{equation}
\frac{d^{2}U}{dz^{2}}+\left(\alpha+\frac{\beta+1}{z}+\frac{\gamma+1}{z-1}\right)\frac{dU}{dz}+\left(\frac{\mu}{z}+\frac{\nu}{z-1}\right)U=0\ ,
\label{eq:Heun_confluente_forma_canonica}
\end{equation}
where $U(z)=\mbox{HeunC}(\alpha,\beta,\gamma,\delta,\eta;z)$ are the confluent Heun functions, with the parameters $\alpha$, $\beta$, $\gamma$, $\delta$ and $\eta$, related to $\mu$ and $\nu$ through the expressions
\begin{equation}
\mu=\frac{1}{2}(\alpha-\beta-\gamma+\alpha\beta-\beta\gamma)-\eta\ ,
\label{eq:mu_Heun_conlfuente_2}
\end{equation}
\begin{equation}
\nu=\frac{1}{2}(\alpha+\beta+\gamma+\alpha\gamma+\beta\gamma)+\delta+\eta\ .
\label{eq:nu_Heun_conlfuente_2}
\end{equation}

Therefore, we have reobtained an analytical solution of the Klein-Gordon equation in the background under consideration that is valid in the region exterior to the event horizon. This means that it includes the regions nearby the event horizon and far from the black hole.

Next, we will consider this radial solution to investigate two interesting phenomena, namely, the resonant frequencies and Hawking radiation.
%
%
\section{Resonant frequencies}\label{Sec.III}
In this section, we use the recently developed technique \cite{AnnPhys.373.28} to compute the resonant frequencies for scalar waves propagating in a GHS dilaton black hole. They are associated with the solution given by Eq.~(\ref{eq:solucao_geral_radial_GHS}), with the boundary conditions that the radial solution should be finite on the exterior event horizon and well behaved at asymptotic infinity. The fact that the solution should be well behaved at asymptotic infinity demands that $R(x)$ must be a polynomial. Indeed, the function $\mbox{HeunC}(\alpha,\beta,\gamma,\delta,\eta;x)$ turns to be a polynomial of degree $n$ if it satisfies the $\delta$-condition, which is given by \cite{MathAnn.33.161,MathComp.76.811,Ronveaux:1995}
\begin{equation}
\frac{\delta}{\alpha}+\frac{\beta+\gamma}{2}+1=-n\ ,
\label{eq:delta_condition}
\end{equation}
where $n=\{0,1,2,\ldots\}$ is a quantum number. This condition will permits us to obtain the desired resonant frequencies.

Thus, substituting Eqs.~(\ref{eq:alpha_radial_HeunC_GHS})-(\ref{eq:delta_radial_HeunC_GHS}) into Eq.~(\ref{eq:delta_condition}), we obtain the expression from which we can determine the resonant frequencies associated to scalar particles in the background under consideration, which is given by
\begin{equation}
\frac{2eQ\omega+r_{h}(\mu_{0}^{2}-2\omega^{2})}{2(\mu_{0}^{2}-\omega^{2})^{\frac{1}{2}}}+i(\omega r_{h}-eQ)=-(n+1)\ .
\label{eq:resonant_frequencies_GHS}
\end{equation}

Equation (\ref{eq:resonant_frequencies_GHS}) can be written as a fourth order equation for $\omega$, namely,
\begin{eqnarray}
&& -2 r_{h}^2 \omega ^4+2 r_{h} (i n+2 e Q+i) \omega ^3 \nonumber\\
&& +(n^2-2 i n e Q+2 n-2 e^2 Q^2-2 i e Q+2 \mu_{0} ^2 r_{h}^2+1) \omega ^2 \nonumber\\
&& -i \mu_{0} ^2 r_{h} (2 n-3 i e Q+2) \omega \nonumber\\
&& -\frac{1}{4} \mu_{0} ^2 (4 n^2-8 i n e Q+8 n-4 e^2 Q^2-8 i e Q+\mu_{0} ^2 r_{h}^2+4) = 0 \ .
\label{eq:fourth_order_omega}
\end{eqnarray}
Note that Eq.~(\ref{eq:fourth_order_omega}) has the general form
\begin{equation}
B_{4}\omega^{4}+B_{3}\omega^{3}+B_{2}\omega^{2}+B_{1}\omega+B_{0}=0\ ,
\label{eq:quartic_equation}
\end{equation}
where $B_{4} \neq 0$. The roots of Eq.~(\ref{eq:fourth_order_omega}) can be determined by using an algebraic manipulation program. As the final expression is so long and no insight is gained by writing it out, we omitted it.

On the other hand, this quantization rule involves a complex number, that is, a frequency (energy) spectrum such that $\omega=\omega_{R}+i\ \omega_{I}$, where $\omega_{R}$ and $\omega_{I}$ are the real and imaginary parts, respectively. Indeed, the main feature of the resonant frequencies corresponds to the decay rate of the oscillation, which is characterized by the imaginary part. Note that the eigenvalues given by Eq.~(\ref{eq:resonant_frequencies_GHS}) are not degenerate, due to the fact that there is no dependence on the eigenvalue $\lambda_{lm}$. We may obtain numerically two values for the resonant frequencies, given by Eq.~(\ref{eq:resonant_frequencies_GHS}), by using the \textit{FindRoot} function in the \textbf{Wolfram Mathematica}$^{\mbox{\tiny\textregistered}}$ \textbf{9}, such that $(\omega-\omega^{(1)}_{n})(\omega-\omega^{(2)}_{n})=0$.

Similar results were obtained in the literature by Ferrari \textit{et al.} \cite{PhysRevD.63.064009} and Shu \textit{et al.} \cite{PhysRevD.70.084046} using an approximate method. In our case we obtain the resonant frequencies directly from the confluent Heun function by using the condition which should be imposed in such a way that this function reduces to a polynomial. In addition, our results includes a discussion about the dependence of the resonant frequencies with the mass of the scalar field, which completes the results in both Refs.~\refcite{PhysRevD.63.064009} and \refcite{PhysRevD.70.084046}.

The resonant frequencies for $n=0$, $e=0.1$, and $\mu_{0}=0.6$ are shown in Tables \ref{tab:resonant_frequencies_GHS_1} and \ref{tab:resonant_frequencies_GHS_2}, where the units $M=1$ were chosen.

\newpage

\begin{table}[htbp]
\tbl{The scalar resonant frequencies $\omega^{(1)}_{n}$ of a GHS dilaton black hole for $e=0.1$ and $\mu_{0}=0.6$. Note that we have focused on the fundamental mode ($n=0$).}
		{\begin{tabular}{ccc}\hline
			$a$  & $\mbox{Re}[\omega^{(1)}_{0}]$ & $\mbox{Im}[\omega^{(1)}_{0}]$ \\\hline
			0.01 & 0.00455 & 0.16642 \\
			0.04 & 0.00911 & 0.16639 \\
			0.09 & 0.01366 & 0.16635 \\
			0.16 & 0.01822 & 0.16629 \\
			0.25 & 0.02277 & 0.16621 \\
			0.36 & 0.02733 & 0.16611 \\
			0.49 & 0.03189 & 0.16599 \\
			0.64 & 0.03645 & 0.16586 \\
			0.81 & 0.04101 & 0.16571 \\
			1.00 & 0.04557 & 0.16554 \\
			1.21 & 0.05014 & 0.16535 \\
			1.44 & 0.05471 & 0.16514 \\
			1.69 & 0.05928 & 0.16491 \\
			1.96 & 0.06385 & 0.16467 \\
			2.25 & 0.06843 & 0.16441 \\\hline
		\end{tabular}
	\label{tab:resonant_frequencies_GHS_1}}
\end{table}

From Table \ref{tab:resonant_frequencies_GHS_1} we see that the real part of the resonant frequencies increase with $a$, while the imaginary part decreases very slowly for fixed values of the charge and mass of the scalar field.

\newpage

\begin{table}[htbp]
\tbl{The scalar resonant frequencies $\omega^{(2)}_{n}$ of a GHS dilaton black hole for $e=0.1$ and $\mu_{0}=0.6$. Note that we have focused on the fundamental mode ($n=0$).}
		{\begin{tabular}{ccc}\hline
			$a$  & $\mbox{Re}[\omega^{(2)}_{0}]$ & $\mbox{Im}[\omega^{(2)}_{0}]$ \\\hline
			0.01 & -0.58659 & 0.04292 \\
			0.04 & -0.58637 & 0.04406 \\
			0.09 & -0.58616 & 0.04522 \\
			0.16 & -0.58595 & 0.04638 \\
			0.25 & -0.58574 & 0.04754 \\
			0.36 & -0.58554 & 0.04872 \\
			0.49 & -0.58535 & 0.04990 \\
			0.64 & -0.58516 & 0.05110 \\
			0.81 & -0.58497 & 0.05230 \\
			1.00 & -0.58479 & 0.05350 \\
			1.21 & -0.58461 & 0.05472 \\
			1.44 & -0.58444 & 0.05594 \\
			1.69 & -0.58427 & 0.05716 \\
			1.96 & -0.58410 & 0.05840 \\
			2.25 & -0.58394 & 0.05964 \\\hline
		\end{tabular}
	\label{tab:resonant_frequencies_GHS_2}}
\end{table}

In Table \ref{tab:resonant_frequencies_GHS_2}, the magnitude of the real part decreases while for the imaginary part it increases with parameter $a$.

\newpage

\begin{figure}[htbp]
	\centering
		\includegraphics[scale=1.00]{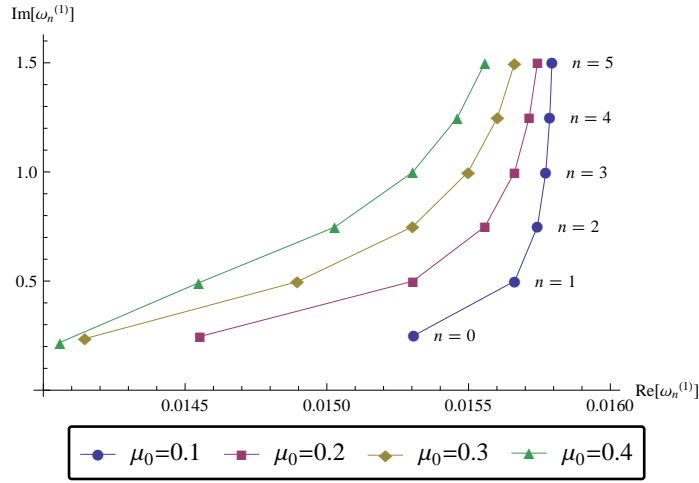}
	\caption{The scalar resonant frequencies of a GHS dilaton black hole as a function of $n$ for $e=0.1$ and $a=0.1$.}
	\label{fig:Fig1_GHS_dilaton}
\end{figure}

\begin{figure}[htbp]
	\centering
		\includegraphics[scale=1.00]{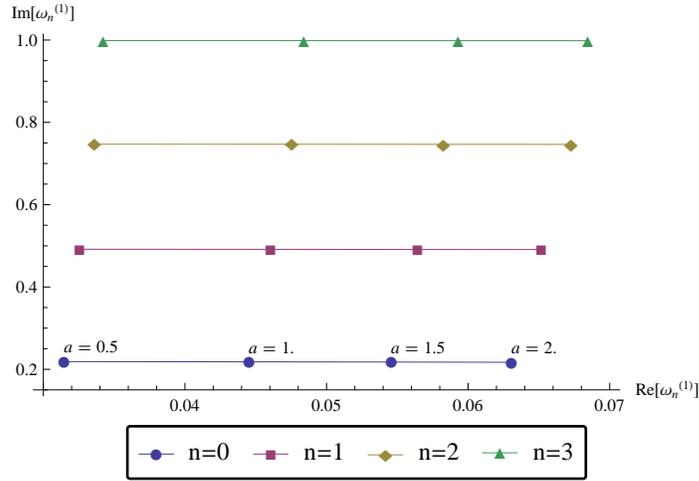}
	\caption{The scalar resonant frequencies of a GHS dilaton black hole as a function of $a$ for $e=0.1$ and $\mu_{0}=0.4$.}
	\label{fig:Fig2_GHS_dilaton}
\end{figure}

\begin{figure}[htbp]
	\centering
		\includegraphics[scale=1.00]{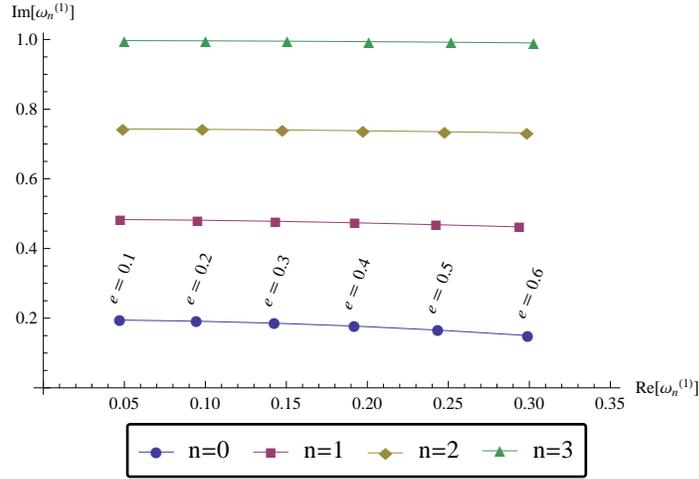}
	\caption{The scalar resonant frequencies of a GHS dilaton black hole as a function of $e$ for $a=1.1$ and $\mu_{0}=0.5$.}
	\label{fig:Fig3_GHS_dilaton}
\end{figure}

\begin{figure}[htbp]
	\centering
		\includegraphics[scale=1.00]{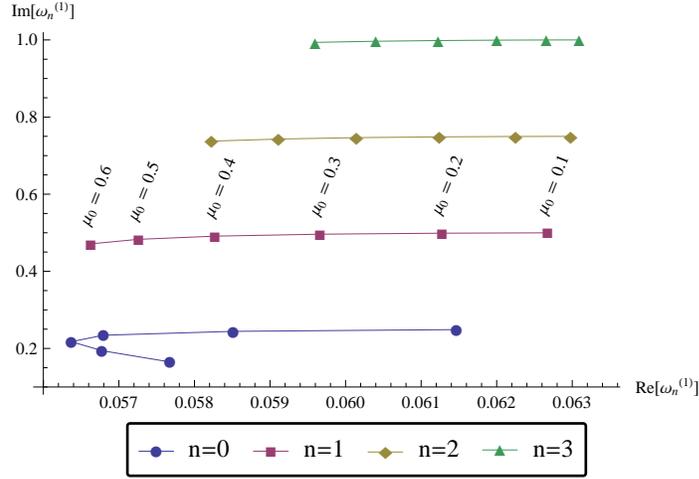}
	\caption{The scalar resonant frequencies of a GHS dilaton black hole as a function of $\mu_{0}$ for $e=0.1$ and $a=1.6$.}
	\label{fig:Fig4_GHS_dilaton}
\end{figure}

The resonant frequencies that we obtained are shown in Figs.~\ref{fig:Fig1_GHS_dilaton}, \ref{fig:Fig2_GHS_dilaton}, \ref{fig:Fig3_GHS_dilaton}, and \ref{fig:Fig4_GHS_dilaton} as a function of $n$, $a$, $e$, and $\mu_{0}$, respectively. The units were chosen as multiples of $M$.

In Fig.~\ref{fig:Fig1_GHS_dilaton} we see that for fixed values of the charge and the parameter $a$, the real part increases with $n$ and decreases with the mass of the field $\mu_{0}$, while the imaginary part is approximately constant with respect to the mass and thus the decay rate of the field is almost constant.

\newpage

From Fig.~\ref{fig:Fig2_GHS_dilaton} we conclude that the real part of the resonant frequencies increases with the values of $n$ and $a$. As to the imaginary part, it increases with $n$, but is approximately constant with respect to the parameter $a$.

\newpage

In this case, shown by Fig.~\ref{fig:Fig3_GHS_dilaton}, where $a$ and $\mu_{0}$ are fixed, the real part of the resonant frequencies increases with the values of $a$ and the charge. As to the imaginary part, it increases with $n$, but is approximately constant with respect to the parameter $a$.

\newpage

In Fig.~\ref{fig:Fig4_GHS_dilaton} we see that, except for the case $n=0$ in which there is anomalous behavior, for all other values of $n$ the real part increases when the mass of the field decreases and $n$ increases, while the imaginary part increases with $n$, but is approximately constant with respect to the mass of the field.

\newpage
%
%
\subsection{Massless case}
In the case where we have a massless scalar field, the expression for the resonant frequencies can be analytically solved for $\omega_{n}$. It is given by
\begin{equation}
\omega_{n}=\frac{eQ}{r_{h}}+i\frac{n+1}{2r_{h}}\ ,
\label{eq:massless_resonant frequencies_GHS}
\end{equation}
where the quantum number $n$ is a positive integer or zero.

\begin{figure}[htbp]
	\centering
		\includegraphics[scale=1.00]{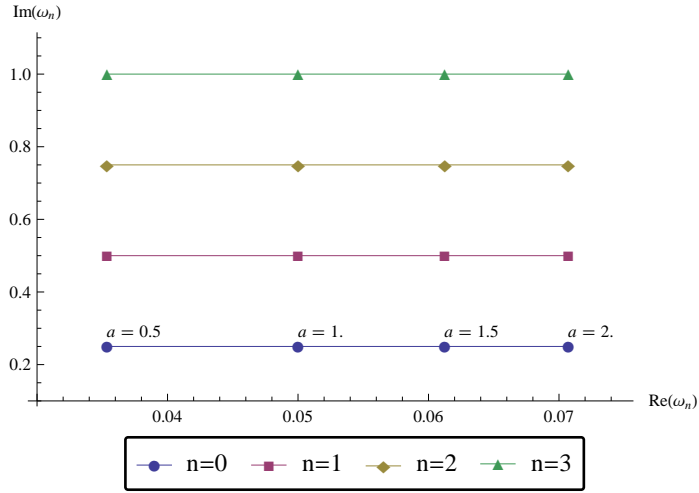}
	\caption{The massless scalar resonant frequencies of a GHS dilaton black hole as a function of $a$ for $e=0.1$.}
	\label{fig:Fig5_GHS_dilaton}
\end{figure}

\begin{figure}[htbp]
	\centering
		\includegraphics[scale=1.00]{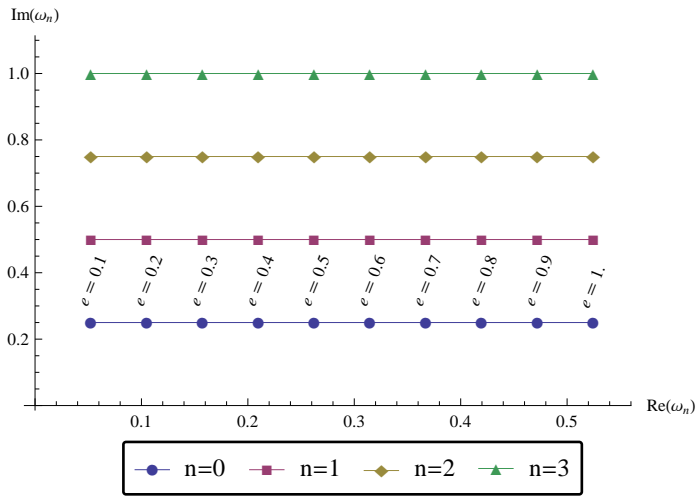}
	\caption{The massless scalar resonant frequencies of a GHS dilaton black hole as a function of $e$ for $a=1.1$.}
	\label{fig:Fig6_GHS_dilaton}
\end{figure}

The field energies given by Eq.~(\ref{eq:massless_resonant frequencies_GHS}) are not degenerate, due to the fact that there is no dependence on the eigenvalue $\lambda_{lm}$. The resonant frequencies for $n=1$, $e=1.1$, and $\mu_{0}=0$ are shown in Table \ref{tab:resonant_frequencies_GHS_3}. In Figs.~\ref{fig:Fig5_GHS_dilaton} and \ref{fig:Fig6_GHS_dilaton}, we present the resonant frequencies as function of $a$ and $e$, respectively. The units were chosen as multiples of $M$.

\begin{table}[htbp]
\tbl{The massless scalar resonant frequencies of a GHS dilaton black hole for $e=1.1$. Note that we have focused on the first excited mode ($n=1$).}
		{\begin{tabular}{ccc}\hline
			$a$  & $\mbox{Re}(\omega_{1})$ & $\mbox{Im}(\omega_{1})$ \\\hline
			0.01 & 0.055 & 0.500 \\
			0.04 & 0.110 & 0.500 \\
			0.09 & 0.165 & 0.500 \\
			0.16 & 0.220 & 0.500 \\
			0.25 & 0.275 & 0.500 \\
			0.36 & 0.330 & 0.500 \\
			0.49 & 0.385 & 0.500 \\
			0.64 & 0.440 & 0.500 \\
			0.81 & 0.495 & 0.500 \\
			1.00 & 0.550 & 0.500 \\
			1.21 & 0.605 & 0.500 \\
			1.44 & 0.660 & 0.500 \\
			1.69 & 0.715 & 0.500 \\
			1.96 & 0.770 & 0.500 \\
			2.25 & 0.825 & 0.500 \\\hline
		\end{tabular}
	\label{tab:resonant_frequencies_GHS_3}}
\end{table}

In Table \ref{tab:resonant_frequencies_GHS_3}, which is related to the massless case, we see that the real part of the resonant frequencies increases with the parameter $a$, while the imaginary part is constant and thus the field will decay with the same rate, independently on the value of $a$.

\newpage

Figure \ref{fig:Fig5_GHS_dilaton} shows that the real part as well as the imaginary part increases with $n$ and $a$, for a fixed value of charge.

\newpage

In Fig.~\ref{fig:Fig6_GHS_dilaton} shown in what follows, the real and imaginary parts increase with $n$ and $e$, for fixed $a$. Thus, the decay rate of the field increases with $n$ and $e$.

\newpage
%
%
\subsection{Special case: Schwarzschild spacetime}
In the case where $a=0$, we get the Schwarzschild black hole. In this scenario, the expression for the resonant frequencies turns into
\begin{equation}
n+1-\frac{r_{h}(\mu_{0}^{2}-2\omega^{2})}{2\sqrt{\mu_{0}^{2}-\omega^{2}}}+i r_{h}\omega=0\ .
\label{eq:resonant frequencies_Sch}
\end{equation}

\begin{figure}[htbp]
	\centering
		\includegraphics[scale=1.00]{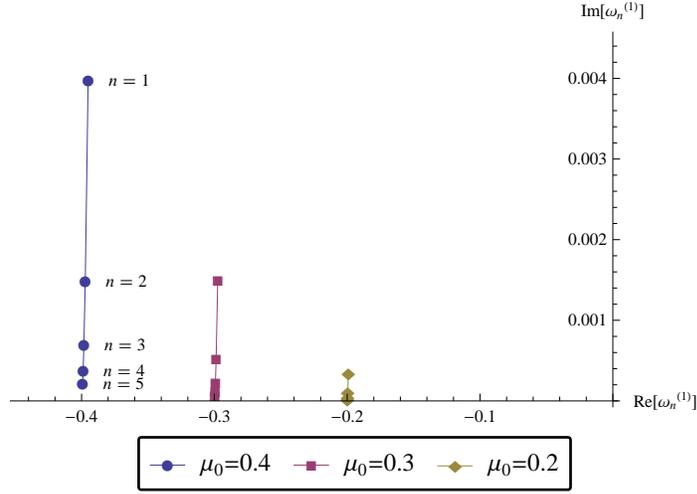}
	\caption{The scalar resonant frequencies $\omega_{1}$ of a Schwarzschild black hole as a function of $\mu_{0}$.}
	\label{fig:Fig7_GHS_dilaton}
\end{figure}

\begin{figure}[htbp]
	\centering
		\includegraphics[scale=1.00]{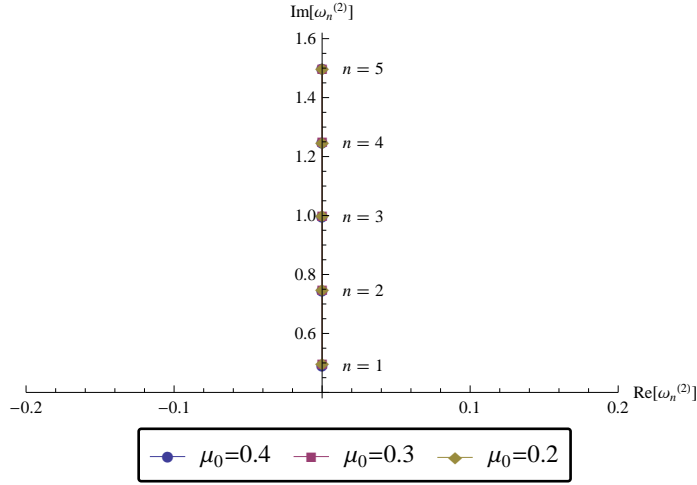}
	\caption{The scalar resonant frequencies $\omega_{2}$ of a Schwarzschild black hole as a function of $\mu_{0}$.}
	\label{fig:Fig8_GHS_dilaton}
\end{figure}

The resonant frequencies associated to massive scalar particles in the Schwarzschild spacetime are shown in Tables \ref{tab:resonant_frequencies_GHS_4}-\ref{tab:resonant_frequencies_GHS_7} and Figs.~\ref{fig:Fig7_GHS_dilaton}-\ref{fig:Fig8_GHS_dilaton}. The units were chosen as multiples of $M$.

\begin{table}[htbp]
\tbl{The scalar resonant frequencies $\omega^{(1)}$ of a Schwarzschild black hole as function of $\mu_{0}$. Note that we have focused on the fundamental mode ($n=0$).}
		{\begin{tabular}{ccc}\hline
			$\mu_{0}$  & $\mbox{Re}[\omega^{(1)}_{0}]$ & $\mbox{Im}[\omega^{(1)}_{0}]$ \\\hline
			0.1 & -0.09957 & 0.00017 \\
			0.2 & -0.19753 & 0.00199 \\
			0.3 &  \ 0.00000 & 0.23611 \\
			0.4 &  \ 0.00000 & 0.21924 \\
			0.5 &  \ 0.00000 & 0.19581 \\
			0.6 &  \ 0.00000 & 0.16643 \\
			0.7 &  \ 0.00000 & 0.13169 \\
			0.8 &  \ 0.00000 & 0.09209 \\
			0.9 &  \ 0.00000 & 0.04808 \\\hline
		\end{tabular}
	\label{tab:resonant_frequencies_GHS_4}}
\end{table}

\begin{table}[htbp]
\tbl{The scalar resonant frequencies $\omega^{(2)}$ of a Schwarzschild black hole as function of $\mu_{0}$. Note that we have focused on the fundamental mode ($n=0$).}
		{\begin{tabular}{ccc}\hline
			$\mu_{0}$  & $\mbox{Re}[\omega^{(2)}_{0}]$ & $\mbox{Im}[\omega^{(2)}_{0}]$ \\\hline
			0.1 &  \ 0.00000 & 0.24965 \\
			0.2 &  \ 0.00000 & 0.24602 \\
			0.3 & -0.29440 & 0.00695 \\
			0.4 & -0.39118 & 0.01538 \\
			0.5 & -0.48852 & 0.02709 \\
			0.6 & -0.58681 & 0.04178 \\
			0.7 & -0.68624 & 0.05916 \\
			0.8 & -0.78691 & 0.07895 \\
			0.9 & -0.88888 & 0.10096 \\\hline
		\end{tabular}
	\label{tab:resonant_frequencies_GHS_5}}
\end{table}

\begin{table}[htbp]
\tbl{The scalar resonant frequencies $\omega^{(1)}$ of a Schwarzschild black hole for $\mu_{0}=0.1$.}
		{\begin{tabular}{ccc}\hline
			$n$  & $\mbox{Re}[\omega^{(1)}_{0}]$ & $\mbox{Im}[\omega^{(1)}_{0}]$ \\\hline
			0 & -0.09957 & 0.00017 \\
			1 & -0.09988 & 0.00002 \\
			2 & -0.09995 & 0.000007 \\
			3 & -0.09997 & 0.000003 \\
			4 & -0.09998 & 0.000001 \\
			5 & -0.09999 & 0.0000009 \\
			6 & -0.09999 & 0.0000005 \\
			7 & -0.09999 & 0.0000003 \\
			8 & -0.09999 & 0.0000002 \\
			9 & -0.10000 & 0.0000001 \\\hline
		\end{tabular}
	\label{tab:resonant_frequencies_GHS_6}}
\end{table}

Tables \ref{tab:resonant_frequencies_GHS_4} and \ref{tab:resonant_frequencies_GHS_5} show the behavior of the real and imaginary parts of $\omega^{(1)}_{0}$ and $\omega^{(2)}_{0}$ as a function of the mass of the field in the case of a Schwarzschild black hole. Note that the decay rate of the field is not uniform.

In Table \ref{tab:resonant_frequencies_GHS_6} the mass was fixed. In this case, the magnitude of the real part of $\omega^{(1)}_{0}$ increases with $n$, while the imaginary part has an irregular behavior.

\newpage

\begin{table}[htbp]
\tbl{The scalar resonant frequencies $\omega^{(2)}$ of a Schwarzschild black hole for $\mu_{0}=0.1$.}
		{\begin{tabular}{ccc}\hline
			$n$  & $\mbox{Re}[\omega^{(2)}_{0}]$ & $\mbox{Im}[\omega^{(2)}_{0}]$ \\\hline
			0 & 0.00000 & 0.24965 \\
			1 & 0.00000 & 0.49995 \\
			2 & 0.00000 & 0.74999 \\
			3 & 0.00000 & 0.99999 \\
			4 & 0.00000 & 1.25000 \\
			5 & 0.00000 & 1.50000 \\
			6 & 0.00000 & 1.75000 \\
			7 & 0.00000 & 2.00000 \\
			8 & 0.00000 & 2.25000 \\
			9 & 0.00000 & 2.50000 \\\hline
		\end{tabular}
	\label{tab:resonant_frequencies_GHS_7}}
\end{table}

In Table \ref{tab:resonant_frequencies_GHS_7}, the mass was fixed. In this case the real part of $\omega^{(2)}_{0}$ is zero, while its imaginary part increases with $n$ and thus the decay rate will increase with $n$.

\newpage

Figure \ref{fig:Fig7_GHS_dilaton} shows how the magnitude of the real pat behaves in terms of $n$ and $\mu_{0}$. It increases when $\mu_{0}$ increases and is constant for different values of $n$. As to the imaginary part, its magnitude increases with $n$ and is almost constant with respect to $\mu_{0}$.

\newpage

In Fig.~\ref{fig:Fig8_GHS_dilaton}, for massive field, the imaginary part of the resonant frequencies increases with the values of $n$.

\newpage

\begin{figure}[htbp]
	\centering
		\includegraphics[scale=1.00]{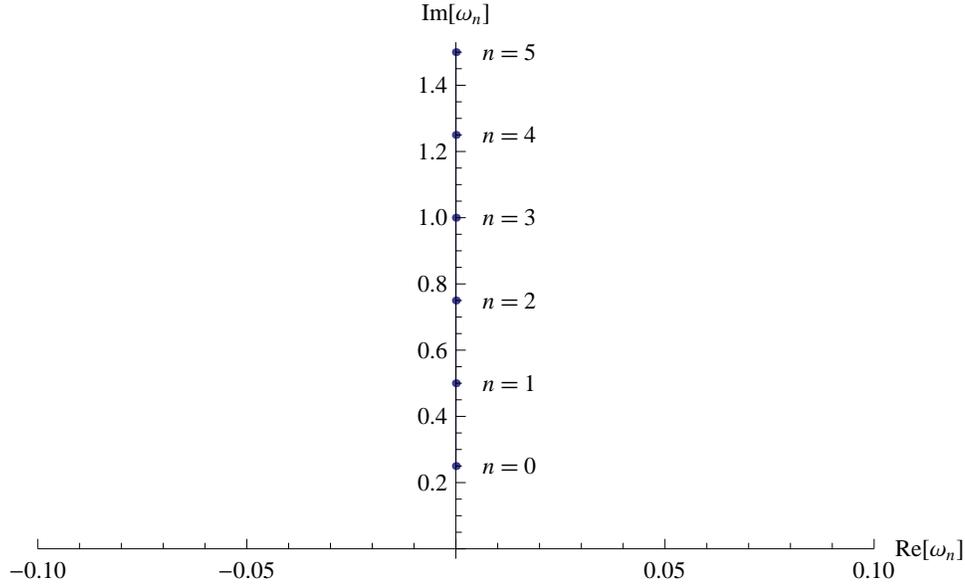}
	\caption{The massless scalar resonant frequencies of a Schwarzschild black hole as a function of $n$.}
	\label{fig:Fig9_GHS_dilaton}
\end{figure}

Now, taking into account a massless scalar field, we find the following expression for the resonant frequencies
\begin{equation}
\omega_{n}=\frac{i(n+1)}{2r_{h}}\ ,
\label{eq:massless_resonant frequencies_Sch}
\end{equation}
which are shown in Table \ref{tab:resonant_frequencies_GHS_8} and Fig.~\ref{fig:Fig9_GHS_dilaton}. The units were chosen as multiples of $M$.

\begin{table}[htbp]
\tbl{The massless scalar resonant frequencies of a Schwarzschild black hole.}
		{\begin{tabular}{ccc}\hline
			$n$  & $\mbox{Re}(\omega_{n})$ & $\mbox{Im}(\omega_{n})$ \\\hline
			0 & 0.00000 & 0.25000 \\
			1 & 0.00000 & 0.50000 \\
			2 & 0.00000 & 0.75000 \\
			3 & 0.00000 & 1.00000 \\
			4 & 0.00000 & 1.25000 \\
			5 & 0.00000 & 1.50000 \\
			6 & 0.00000 & 1.75000 \\
			7 & 0.00000 & 2.00000 \\
			8 & 0.00000 & 2.25000 \\
			9 & 0.00000 & 2.50000 \\\hline
		\end{tabular}
	\label{tab:resonant_frequencies_GHS_8}}
\end{table}

These results of Table \ref{tab:resonant_frequencies_GHS_8} show us that for the massless field in Schwarzschild, the real part is zero and the imaginary part increases with $n$, which means that the decay rate increases with this number.

\newpage

Figure \ref{fig:Fig9_GHS_dilaton} shows that the imaginary part of the resonant frequencies increases with $n$. This is the same qualitative behavior of the massive field whose results are given in Fig.~\ref{fig:Fig8_GHS_dilaton}.

It is worth commenting that the results obtained in the case of GHS dilaton black hole seem to be qualitatively in accordance with the ones obtained by Ferrari \textit{et al.} \cite{PhysRevD.63.064009}. The same to be happen with respect to the results concerning the Schwarzschild case, as compared with recent results in the literature \cite{PhysRevD.69.044004}, taking into consideration the qualitative aspects. In which concerns the values obtained, they are different, which is expected due to the fact that we have used different boundary conditions as compared to the ones used in the literature.

All these results can be obtained in the case of stationary black holes, in principle. To do this we have to use the exact solutions of the Klein-Gordon equation due to the fact that the method is based on the exact solution of this equation. In the literature, we have some results in the Kerr black hole spacetime with the use of the approximate methods \cite{PhysRevD.84.044046,PhysLettB.715.348}.

\newpage
%
%
\section{Some aspects of Hawking radiation}\label{Sec.IV}
In this section, we will study and discuss some aspects of the black body radiation emitted by a GHS dilaton black hole. In order to do this, we need to consider the radial solution near the exterior event horizon, that is, to analyze the radial solution when $r \rightarrow r_{h}$ which implies that $x \rightarrow 0$. Similar studies were done in Ref.~\refcite{CanJPhys.87.349}, which obtained the Hawking radiation spectrum and the relative scattering probability. Additionally we discussed the flux of particles and the free energy and how it depends on the dilaton charge.

Firstly, let us consider the expansion in power series of the confluent Heun function with respect to the independent variable $x$, in a neighborhood of the regular singular point $x=0$, which can be written as \cite{Ronveaux:1995}
\begin{eqnarray}
\mbox{HeunC}(\alpha,\beta,\gamma,\delta,\eta;x) & = & 1+\frac{1}{2}\frac{(-\alpha\beta+\beta\gamma+2\eta-\alpha+\beta+\gamma)}{(\beta+1)}x\nonumber\\
& + & \frac{1}{8}\frac{1}{(\beta+1)(\beta+2)}(\alpha^{2}\beta^{2}-2\alpha\beta^{2}\gamma+\beta^{2}\gamma^{2}\nonumber\\
& - & 4\eta\alpha\beta+4\eta\beta\gamma+4\alpha^{2}\beta-2\alpha\beta^{2}-6\alpha\beta\gamma\nonumber\\
& + & 4\beta^{2}\gamma+4\beta\gamma^{2}+4\eta^{2}-8\eta\alpha+8\eta\beta+8\eta\gamma\nonumber\\
& + & 3\alpha^{2}-4\alpha\beta-4\alpha\gamma+3\beta^{2}+4\beta\delta\nonumber\\
& + & 10\beta\gamma+3\gamma^{2}+8\eta+4\beta+4\delta+4\gamma)x^2+...\ .
\label{eq:serie_HeunC_todo_x}
\end{eqnarray}

Thus, in this limit, the radial solution given by Eq.~(\ref{eq:solucao_geral_radial_GHS}) becomes
\begin{equation}
R(r) \sim C_{1}\ (r-r_{h})^{\beta/2}+C_{2}\ (r-r_{h})^{-\beta/2}\ ,
\label{eq:exp_0_solucao_geral_radial_GHS}
\end{equation}
where we have only considered contributions of the first term in the expansion, and all constants are included in $C_{1}$ and $C_{2}$. Then, considering the solution of the time dependence, near the exterior event horizon $r_{h}$ of the GHS dilaton black hole, we can write
\begin{equation}
\Psi=\mbox{e}^{-i \omega t}(r-r_{h})^{\pm\beta/2}\ .
\label{eq:sol_onda_radial_GHS}
\end{equation}
Taking into account Eq.~(\ref{eq:beta_radial_HeunC_GHS}), the parameter $\beta$ can be rewritten as
\begin{equation}
\frac{\beta}{2}=i(\omega r_{h}-eQ)=\frac{i}{2\kappa_{h}}(\omega-\omega_{h})\ ,
\label{eq:beta/2_solucao_geral_radial_GHS}
\end{equation}
where 
\begin{equation}
\omega_{h}=e\Phi_{h}\ ,
\label{eq:omega0}
\end{equation}
with $\Phi_{h}$ being the electric potential nearby the exterior event horizon ans is given by
\begin{equation}
\Phi_{h}(M,a)=\frac{\partial M}{\partial Q}\biggr|_{S_{h}}=\frac{\sqrt{aM}}{r_{h}}\ .
\label{eq:pot_ele_a_GHS}
\end{equation}
The parameter $\kappa_{h}$ is the gravitational acceleration on the background exterior event horizon surface, and can be written as
\begin{equation}
\kappa_{h}=\frac{1}{2r_{h}}\ .
\label{eq:gravitational_acceleration_GHS}
\end{equation}

Therefore, in the GHS dilaton black hole exterior horizon surface, the ingoing and outgoing wave solutions are given by
\begin{equation}
\Psi_{in}=\mbox{e}^{-i \omega t}(r-r_{h})^{-\frac{i}{2\kappa_{h}}(\omega-\omega_{h})}\ ,
\label{eq:sol_in_1_GHS}
\end{equation}
\begin{equation}
\Psi_{out}(r>r_{h})=\mbox{e}^{-i \omega t}(r-r_{h})^{\frac{i}{2\kappa_{h}}(\omega-\omega_{h})}\ .
\label{eq:sol_out_2_GHS}
\end{equation}

Next, we obtain the expression for the relative scattering probability of the scalar wave, at the exterior event horizon surface, which is
\begin{equation}
\Gamma_{h}(\omega)=\left|\frac{\Psi_{out}(r>r_{h})}{\Psi_{out}(r<r_{h})}\right|^{2}=\mbox{e}^{-\frac{2\pi}{\kappa_{h}}(\omega-\omega_{h})}=\mbox{e}^{-\beta_{h}(\omega-\omega_{h})}\ ,
\label{eq:taxa_refl_GHS}
\end{equation}
where the thermodynamic quantity $\beta_{h}$ is given by\begin{equation}
\beta_{h}=\frac{1}{k_{B}T_{h}}\ ,
\label{eq:beta_GHS}
\end{equation}
with $T_{h}$ being the Hawking temperature, which is related to the gravitational acceleration by the expression
\begin{equation}
T_{h}=\frac{\partial M}{\partial S_{h}}\biggr|_{Q}=\frac{\kappa_{h}}{2\pi}\ ,
\label{eq:Hawking_temperature_GHS}
\end{equation}
which is the same of the Schwarzschild black hole.

Thus, by using the method developed by Damour-Ruffini-Sannan \cite{PhysRevD.14.332,GenRelativGravit.20.239}, we get the resulting Hawking radiation spectrum of charged massive scalar particles (mean number of particles emitted), which is written as
\begin{equation}
\bar{N}_{\omega}=\frac{\Gamma_{h}}{1-\Gamma_{h}}=\frac{1}{\mbox{e}^{\beta_{h}(\omega-\omega_{h})}-1}\ .
\label{eq:espectro_rad_GHS_2}
\end{equation}
Equation (\ref{eq:espectro_rad_GHS_2}) give us the black body spectrum described by scalar particles which are emitted from the GHS dilaton black hole. It is worth calling attention to the fact that the results obtained until Eq.~(\ref{eq:espectro_rad_GHS_2}) are similar the ones obtained in the literature \cite{CanJPhys.87.349}. From now on we will present the new results that follows.

In the limit where $\omega$ is a continuous variable, this yields a total rate of particle emission, which is expressed as
\begin{equation}
\frac{dN}{dt}=\int_{0}^{\infty}\frac{\bar{N}_{\omega}}{2\pi}\ d\omega=\frac{1}{\beta_{h}}\ln\left(\frac{1}{1-\mbox{e}^{\beta_{h}\omega_{h}}}\right)\ .
\label{eq:total_rate_GHS}
\end{equation}

The mass loss rate can be calculated using the following relation
\begin{equation}
\frac{dM}{dt}=-\frac{1}{2\pi}\int_{0}^{\infty}\bar{N}_{\omega}\ \omega\ d\omega=-\frac{1}{2\pi}\frac{1}{\beta_{h}^{2}}\mbox{Li}_{2}(\mbox{e}^{\beta_{h}\omega_{h}})\ ,
\label{eq:Mass_loss_GHS}
\end{equation}
where $\mbox{Li}_{n}(z)$ is the poly-logarithm function with $n$ running from 1 to $\infty$.

The flux of scalar particles, $\Phi$, at infinity, i.e., far from the GHS dilaton black hole, is given by
\begin{equation}
\Phi=\left|\frac{dM}{dt}\right|=\frac{1}{2\pi}\frac{1}{\beta_{h}^{2}}\mbox{Li}_{2}(\mbox{e}^{\beta_{h}\omega_{h}})\ .
\label{eq:Flux_GHS}
\end{equation}

\begin{figure}[htbp]
	\centering
		\includegraphics[scale=1.35]{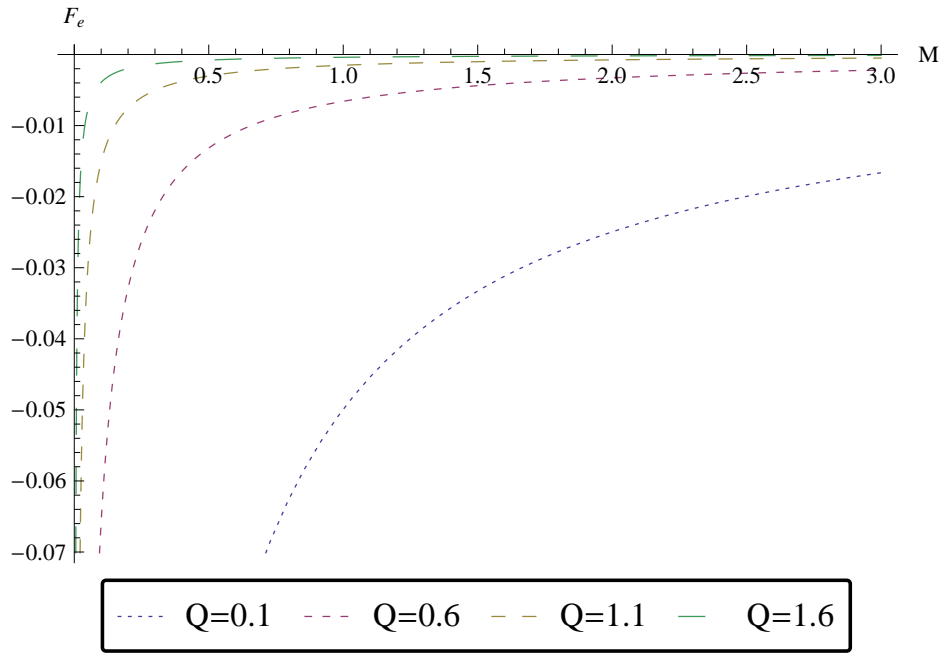}
	\caption{Free energy of the scalar particle in a GHS dilaton black hole as a function of $M$ for $e=-0.1$.}
	\label{fig:Fig10_GHS_dilaton}
\end{figure}

Finally, according to the canonical assembly theory \cite{CommunTheorPhys.52.189}, considering that the frequency (energy) is continuous, the free energy of the scalar particle, $F_{e}$, can be expressed as
\begin{equation}
F_{e}=-\int_{0}^{\infty}\frac{\Gamma_{h}(\omega)}{\mbox{e}^{\beta_{h}(\omega-\omega_{h})}-1}\ d\omega=\frac{\mbox{e}^{\beta_{h}\omega_{h}}+\ln(1-\mbox{e}^{\beta_{h}\omega_{h}})}{\beta_{h}}\ .
\label{eq:free_energy_GHS}
\end{equation}
In Fig.~\ref{fig:Fig10_GHS_dilaton} is shown the free energy of the scalar particle as a function of $M$, for different values of $Q$. Note that the free energy of the scalar particle will be a real number only if the magnetic charge $Q$ or the charge of the scalar particle $e$ is negative.

This result given by Fig.~\ref{fig:Fig10_GHS_dilaton} tell us that the free energy increases with the dilaton charge as well as with the mass of the GHS dilaton black hole.

\newpage
%
%
\section{Summary and discussion}\label{Sec.V}
In this work, we have studied the interaction between scalar fields and the GHS dilaton black hole, and used the Klein-Gordon equation to investigate some physical process which correspond to the resonant frequencies and Hawking radiation.

These solutions confirm and extend the ones known in the literature, in the sense that now we have analytic solutions that are valid for a spacetime outside the exterior event horizon, which means, in the region between the event horizon and infinity, differently from the results obtained in \cite{NuclearPhysicsB.899.37,ClassQuantumGrav.22.533,CommunTheorPhys.52.189} which are valid only in asymptotic regions, namely, very close to the interior event horizon and far from the black hole. This approach has the advantage that it is not necessary to introduce any coordinate system, as for example, the particular ones used in \cite{IntJTheorPhys.52.1682}.

We have imposed boundary conditions to the radial solution of the Klein-Gordon equation in order to analyze the resonant frequencies and then obtained a general expression for these oscillations. By using a numerical approach, we present some values for the resonant frequencies as functions of the involved parameters. We also calculated and discussed the massless resonant frequencies.

Considering the resonant frequencies we can conclude that for the fundamental mode, taking into account a massive field, the real and imaginary parts of the resonant frequencies $\omega^{(1)}_{0}$ of the quasispectrum decreases when the parameter $a$, which codifies the presence of the dilaton, decreases, as shown in Table \ref{tab:resonant_frequencies_GHS_1}. For fixed value of $a$, the behaviors of the real and imaginary parts of the resonant frequencies, for different modes, is shown in Fig.~\ref{fig:Fig1_GHS_dilaton}. Similarly, in Table \ref{tab:resonant_frequencies_GHS_2} and Fig.~\ref{fig:Fig2_GHS_dilaton}, the behaviors of $\omega^{(2)}_{0}$ and $\omega^{(2)}_{n}$ ($n=\{0,1,2,3\}$) are shown, respectively.

Figure \ref{fig:Fig3_GHS_dilaton} shows that for fixed values of $a$ and $\mu_{0}$, for the modes $n=\{0,1,2,3\}$, the real part of $\omega^{(1)}_{n}$ increases with the value of the charge $e$, while the imaginary part decreases slightly. Fig.~\ref{fig:Fig4_GHS_dilaton} shows the behavior of the resonant frequencies, but now as a function of $\mu_{0}$, for different modes.

For the fundamental mode, the behavior is not regular, while for the others modes, namely, $n=\{1,2,3\}$, the real part of the resonant frequencies increases with the decreasing of $\mu_{0}$, while the imaginary part is approximately constant.

A massless scalar field was also considered, for which behavior of the resonant frequencies is shown in Table \ref{tab:resonant_frequencies_GHS_3} for the first excited mode. Note that the real part increases, while the imaginary part is constant, with the increasing of the parameter $a$. This behavior is reconfirmed in Figs.~\ref{fig:Fig5_GHS_dilaton} and \ref{fig:Fig6_GHS_dilaton}, when different modes are considered, for fixed values of the charge $e$ and the parameter $a$, respectively.

Table \ref{tab:resonant_frequencies_GHS_4} shows the behavior of the resonant frequencies for different values of $\mu_{0}$, in the Schwarzschild case, for the fundamental mode. Comparing the results taking the values of $\mu_{0}$ from 0.1 to 0.6 of this Table with the results shown in Fig.~\ref{fig:Fig3_GHS_dilaton} for $n=0$, we can see how is the effect of the presence of the dilaton, in this particular case. Similar results should happens for all others configurations, as should be expected. Similarly, comparing Fig.~\ref{fig:Fig8_GHS_dilaton} with Fig.~\ref{fig:Fig4_GHS_dilaton} and Fig.~\ref{fig:Fig9_GHS_dilaton} with Fig.~\ref{fig:Fig5_GHS_dilaton}, we can see how is the role played by the dilaton.

Comparing Tables \ref{tab:resonant_frequencies_GHS_1} and \ref{tab:resonant_frequencies_GHS_2} with Tables \ref{tab:resonant_frequencies_GHS_4}, \ref{tab:resonant_frequencies_GHS_5}, \ref{tab:resonant_frequencies_GHS_6} and \ref{tab:resonant_frequencies_GHS_7} for the massive case, we can see the influence of the dilaton. Also comparing Tables \ref{tab:resonant_frequencies_GHS_3} and \ref{tab:resonant_frequencies_GHS_8}, for the massless case, the role made by the dilaton becomes evident.

The Hawking radiation spectrum was obtained from the asymptotic behavior of the radial solution at the exterior event horizon, where we have used the expansion in power series of the confluent Heun function.

Finally, we obtained the Hawking flux of scalar particles from the GHS dilaton black hole given by Eq.~(\ref{eq:Flux_GHS}), from which we can conclude that this quantity depends on the parameter which codifies the presence of the dilaton through its influence on the Hawking temperature. Also, the free energy depends on the charge $Q$ and mass of the black hole $M$, and consequently on the presence of the dilaton field.

It is worth calling attention to the fact that the determination of the resonant frequencies and flux of particles emitted as a Hawking radiation can be done for the case of a more realistic black hole, as for example, in Kerr black holes, by using the same method that we have adopted. Certainly, this will be more difficult due the fact that the solution of the Klein-Gordon equation will be more complicated and not necessarily has an explicitly polynomial form.
%
%
\section*{Acknowledgments}
H. S. V. is funded by the CNPq through the research Project (150640/2018-8). V. B. B. is partially supported by the CNPq through the research Project (305835/2016-5). M. S. C. is partially supported by the CNPq through the research Project (312251/2015-7).
%
%

%
%

\begin{thebibliography}{99}
%
\bibitem{PhysRevLett.116.221101} B. P. Abbott \textsl{et al.} [Ligo Scientific and Virgo Collaborations], Phys. Rev. Lett. \textbf{116}, 221101 (2016).
\bibitem{AstrophysJLett.875.L1} The Event Horizon Collaboration, Astrophys. J. Lett. \textbf{875}, L1 (2019).
\bibitem{NuclPhysB.207.337} G. W. Gibbons, Nucl. Phys. B \textbf{207}, 337 (1982).
\bibitem{AnnPhys.172.304} R. C. Myers and M. J. Perry, Ann. Phys. (NY) \textbf{172}, 304 (1986).
\bibitem{NuclPhysB.289.701} R. C. Myers, Nucl. Phys. B \textbf{289}, 701 (1987).
\bibitem{NuclPhysB.309.552} H. J. Vega and N. Sanchez, Nucl. Phys. B \textbf{309}, 552 (1988).
\bibitem{NuclPhysB.298.741} G. W. Gibbons and K. Maeda, Nucl. Phys. B \textbf{298}, 741 (1988).
\bibitem{PhysRevD.43.3140} D. Garfinkle, G. T. Horowitz and A. Strominger, Phys. Rev. D \textbf{43}, 3140 (1991); Erratum Phys. Rev. D \textbf{45}, 3888 (1992).
\bibitem{PhysRevLett.69.1006} A. Sen, Phys. Rev. Lett. \textbf{69}, 1006 (1992).
\bibitem{CanJPhys.87.349} I. Sakali and A. Al-Badawi, Can. J. Phys. \textbf{87}, 349 (2015).
\bibitem{arXiv:1807.09135v1} H. S. Vieira, V. B. Bezerra, C. R. Muniz, M. S. Cunha and M. O. Tahim, arXiv:1807.09135v1 \textbf{[gr-qc]} (2018).
\bibitem{AdvHighEnergyPhys.2019.5769564} M. A. Dariescu, C. Dariescu and C. Stelea, Adv. High Energy Phys. \textbf{2019}, 5769564 (2019).
\bibitem{ChinPhysB.19.090401} M. J. Lan, G. Chen and Y. W. Han, Chin. Phys. B \textbf{19}, 090401 (2010).
\bibitem{NuovoCimento.122.904} X. M. Liu, L. C. Zhang and R, Zhao, Nuovo Cimento \textbf{122}, 904 (2007).
\bibitem{CommunTheorPhys.52.184} C. Y. Wang and Y. X. Gui, Commun. Theor. Phys. \textbf{52}, 189 (2009).
\bibitem{PhysRevD.63.064009} V. Ferrari, M. Pauri and F. Piazza, Phys. Rev. D \textbf{63}, 064009 (2001).
\bibitem{PhysRevD.70.084046} F. W. Shu and Y. G. Shen, Phys. Rev. D \textbf{70}, 084046 (2004).
\bibitem{ClassQuantumGrav.22.1129} S. Chen and J. Jing, Class. Quantum Grav. \textbf{22}, 1129 (2005).
\bibitem{AdvHighEnergyPhys.2015.739153} I. Sakali and G. Tokgoz, Adv. High Energy Phys. \textbf{2015}, 739153 (2015).
\bibitem{PhysRevD.81.104042} S. W. Wei, Y. X. Liu, K. Yang and Y. Zhang, Phys. Rev. D \textbf{81}, 104042 (2010).
\bibitem{Detweiler:1979} S. L. Detweiler, \textit{Sources of gravitational radiation}, edited by L. Smart (Cambridge University Press, Cambridge, 1979).
\bibitem{PhysLettB.761.53} S. Hod, Phys. Lett. B \textbf{761}, 53 (2016).
\bibitem{EurPhysJPlus.132.324} B. Toshmatov and Z. Stuchlík, Eur. Phys. J. Plus \textbf{132}, 324 (2017).
\bibitem{Nature.248.30} S. W. Hawking, Nature \textbf{248}, 30 (1974).
\bibitem{CommunMathPhys.43.199} S. W. Hawking, Commun. Math. Phys. \textbf{43}, 199 (1975).
\bibitem{AnnPhys.373.28} H. S. Vieira and V. B. Bezerra, Ann. Phys. (NY) \textbf{373}, 28 (2016).
\bibitem{PhysRevD.94.084040} I. Sakalli, Phys. Rev. D \textbf{94}, 084040 (2016).
\bibitem{NuclearPhysicsB.899.37} C. Y. Zhang, S. J. Zhang and B. Wang, Nuclear Physics B \textbf{899}, 37 (2015).
\bibitem{ClassQuantumGrav.22.533} S. Chen and J. Jing, Class. Quantum Grav. \textbf{22}, 533 (2005).
\bibitem{AstrophysSpaceSci.333.369} K. Lin, Astrophys. Space Sci. \textbf{333}, 369 (2011).
\bibitem{IntJTheorPhys.49.2786} K. Lin, Int. J. Theor. Phys. \textbf{49}, 2786 (2010).
\bibitem{IntJTheorPhys.52.1474} H. Liao, J. Chen and Y. Wang, Int. J. Theor. Phys. \textbf{52}, 1474 (2013).
\bibitem{ClassQuantumGrav.31.045003} V. B. Bezerra, H. S. Vieira and A. A. Costa, Class. Quantum Grav. \textbf{31}, 045003 (2014).
\bibitem{AnnPhys.350.14} H. S. Vieira, V. B. Bezerra and C. R. Muniz, Ann. Phys. (NY) \textbf{350}, 14 (2014).
\bibitem{JCAP01(2018)006} C. R. Muniz, M. O. Tahim, M. S. Cunha and H. S. Vieira, JCAP \textbf{001}, 006 (2018).
\bibitem{JPhysAMathTheor.43.035203} P. P. Fiziev, J. Phys. A: Math. Theor. \textbf{43}, 035203 (2010).
\bibitem{MathAnn.33.161} K. Heun, Math. Ann. \textbf{33}, 161 (1889).
\bibitem{MathComp.76.811} R. Maier, Math. Comp. \textbf{76}, 811 (2007).
\bibitem{Ronveaux:1995} A. Ronveaux, \textit{Heun's differential equations}, (Oxford University Press, New York, 1995).
\bibitem{PhysRevD.69.044004} V. Cardoso, J. P. S. Lemos and S. Yoshida, Phys. Rev. D \textbf{69}, 044004 (2004).
\bibitem{PhysRevD.84.044046} S. Hod, Phys. Rev. D \textbf{84}, 044046 (2011).
\bibitem{PhysLettB.715.348} S. Hod, Phys. Lett. B \textbf{715}, 348 (2012).
\bibitem{PhysRevD.14.332} T. Damour and R. Ruffini, Phys. Rev. D \textbf{14}, 332 (1976).
\bibitem{GenRelativGravit.20.239} S. Sannan, Gen. Relativ. Gravit. \textbf{20}, 239 (1988).
\bibitem{CommunTheorPhys.52.189} C. Y. Wang and Y. X. Gui, Commun. Theor. Phys. \textbf{52}, 189 (2009).
\bibitem{IntJTheorPhys.52.1682} X. G. Lan, Int. J. Theor. Phys. \textbf{52}, 1682 (2013).
\end{thebibliography}
\end{document}